\documentclass[a4paper,11pt]{article}
\usepackage{amsmath,amssymb,amsbsy,amsfonts,latexsym,amsthm}
\usepackage{graphicx}
\newcommand{\be}{\begin{equation}}    
\newcommand{\ee}{\end{equation}}
\newcommand{\ba}{\begin{eqnarray}}
\newcommand{\ea}{\end{eqnarray}}
\newcommand{\bv}{\boldsymbol}

\newcommand{\sizedef}{\headheight=0pt \topmargin=-1.5cm \headsep=1.5cm              
        \oddsidemargin=-0.5cm \evensidemargin=-0.5cm  
        \textheight=22truecm \textwidth=16.5truecm \setlength{\columnsep}{20pt}}
\sizedef

\begin{document}

\bibliographystyle{prsty}

\title{Autonomous perturbations of LISA orbits}
\author{G. Pucacco\thanks{e-mail: pucacco@roma2.infn.it},  M. Bassan %\thanks{e-mail: bassan@roma2.infn.it} \\
Dipartimento di Fisica -- Universit\`a di Roma ``Tor Vergata'' \\ 
\& \\
INFN -- Sezione di Roma Tor Vergata\\
\\
M. Visco %\thanks{e-mail: massimo.visco@ifsi-roma.inaf.it} \\
IFSI-INAF \& INFN -- Sezione di Roma Tor Vergata\\
}
\date{}
\maketitle

\begin{abstract}
We investigate autonomous perturbations on the orbits of LISA, namely the effects produced by gravitational fields that can be expressed only in terms of the position, but {\it not of time} in the Hill frame. This first step in the study of the LISA orbits has been the subject of recent papers which implement analytical techniques based on a ``post-epicyclic'' approximation in the Hill frame to find optimal unperturbed orbits. The natural step forward is to analyze the perturbations to purely Keplerian orbits. In the present work a particular emphasis is put on the tidal field of the Earth assumed to be stationary in the Hill frame. Other relevant classes of autonomous perturbations are those given by the corrections to the Solar field responsible for a slow precession and a global stationary field, associated to sources like the interplanetary dust or a local dark matter component. The inclusion of simple linear contributions in the expansion of these fields produces secular solutions that can be compared with the measurements and possibly used to evaluate some morphological property of the perturbing components.
\end{abstract}

% \vfill\eject

\section{Introduction}
Recent papers have revived the analytical study of the orbits of the three spacecrafts of the LISA `constellation' \cite{cqg1,cqg2,sw}.  Although the topic has been explored and analyzed in some detail with the usual numerical methods and many perturbing effects were already considered and dealt with \cite{hu,pk}, we think it can be useful to pursue an analytical approach, that can be of help in gaining insight on the relevance and hierarchy of the various perturbing effects.

 The main results obtained so far include the possibility of realizing first-order stable solutions of arbitrary shape in both `standard'  planes at angles near $\pm60^{\circ}$ with respect to the ecliptic and the optimization of the configuration which produces a peak-to-peak variation in arm length of about 40000 km and a maximum rate of change of arm length
of about 4 m/s. These figures represent the least possible amount of `flexing' of the arms of the nominal equilateral triangular configuration and refer to a purely Keplerian motion in the field of the Sun only, on suitably phase-shifted Earth-like orbits. They can be obtained either by a trial/error procedure based on the direct computation of the exact Keplerian orbits \cite{hu} or by a perturbative analytical study. It can be shown \cite{cqg2} that, due to the small eccentricity and inclination of the orbits, a series expansion up to second order in a suitable perturbation parameter differs from the exact Keplerian solution by less than $0.03 \%$ making the method of analytical series expansion a viable route to study more complex aspects of the motion of LISA.
	
	There are in fact several sources of perturbations affecting the motion of the three spacecrafts on the unperturbed elliptic orbits. It is easy to convince ourselves that all of them are weak enough to produce very small effects over the time of the mission. However, in view of the need of an accurate knowledge of distances and Doppler shifts in the implementation of the Time Delay Interferometry, knowledge, analysis and, possibly, mitigation of all perturbative effects are a necessary task. A preliminary study can be profitably done with those analytical techniques mentioned above, in parallel with more accurate but painstaking numerical simulations \cite{hu,pk}.
	
The most useful setting in which to perform the analytical study is that based on the Hill system, recalled in sect.3. This is a frame rotating with a constant angular velocity set by the mean motion along a reference orbit. The dynamics in this frame are determined by a tidal approximation of the gravitational potential of the Sun and other sources. The {\it far tide} approximation \cite{bp} consists in taking only the first order term in the field expansion around the origin of the frame. In order to be consistent with the above mentioned series expansion of the Keplerian orbits, it is sufficient to truncate the expansion to the second (`octupolar') order of the field of the Sun. The same procedure can then be applied to the Earth, planets, etc.
	
These perturbations can be classified as {\it autonomous} and {\it non-autonomous}. In the first class fall the fields that can be expressed only in terms of the position (and, possibly, of velocity), but {\it not of time} in the Hill frame.
In this paper we shall focus on the autonomous perturbation effects on LISA orbits. The most relevant example in this class is the field of the Earth, that in the Hill frame is practically at rest. 
The equations of motion can be cast in the form of a system of oscillators subject to a linear vector potential (due to the Coriolis acceleration) and coupled through higher-order terms. The linear system can be solved exactly, the nonlinear couplings are then included perturbatively. In sect.4 we  investigate the configuration, i.e. the phase of each LISA spacecraft orbit, at the closest approach to the Earth, in order to see the effects on all the flexing and Doppler shift indicators, in particular when the time base-line of the mission is longer than usually assumed. 

The effects of global stationary fields can also be studied by their series expansion in the Hill frame. The linear part already provides useful information. A first interesting example, that we address in sect.5 is given by the corrections to the gravitational potential of the Sun, treated as a classical spherical mass. The general relativistic correction in the weak field limit, responsible for the classical relativistic perihelion advance, can be obtained by adding a term scaling as the inverse cube of the distance and parametrized by the Schwarzschild radius. Another interesting correction term is provided by the intrinsic quadrupole moment of the Sun. With the current values of the parameters, the effects produced by both these terms are probably within the detection capabilities of LISA. 
In sect.6 we discuss the tiny effects of the interplanetary dust or a local dark matter component that can in principle be traced in the spectrum of the relative motion between the spacecrafts and can be used to evaluate some morphological property of the perturbing component.

In general, the perturbative effects due to all other bodies in the Solar System have to be expressed by fields explicitly dependent on time and therefore fall in the non-autonomous class. In sect.7 we conclude with some perspective on the study of this second class of perturbations.

\section{Unperturbed Keplerian orbits}

We briefly recall the main results,  obtained by other authors,  that are useful for our work.
We begin by writing  the unperturbed motion of the three spacecrafts $SC_{k}, \; k=1,2,3$,  on Keplerian orbits in the field of the Sun alone. In solar barycentric coordinates, let us call $X_{k},Y_{k},Z_{k}$ the positions of the $SC_{k}$ along the three orbits. They are given by \cite{cqg1}:

\begin{eqnarray}\label{eq:kepler}
X_{k}&=&X_{1} (\psi_{k}) \cos \frac{2 \pi}3 (k-1) - Y_{1} (\psi_{k}) \sin \frac{2 \pi}3 (k-1), \nonumber \\
Y_{k}&=&X_{1} (\psi_{k}) \sin \frac{2 \pi}3 (k-1) + Y_{1} (\psi_{k}) \cos \frac{2 \pi}3 (k-1), \\
Z_{k}&=&Z_{1} (\psi_{k}), \nonumber
\end{eqnarray}
where
$\psi_{k}$ is the {\it eccentric anomaly} of $SC_{k}$ (see below, Eq.\ref{eccentric}). We then introduce the natural series expansion parameter

\be\label{alfa}
\alpha= \frac{\ell}{2 R} \simeq \frac1{60},\ee
where
 $\ell = 5 \times 10^6 \; {\rm km}$
is the nominal length of an arm (the side of the LISA triangle)
and
$ R = 1.5 \times 10^8 \; {\rm km} = 1 {\rm UA}$
is the semi-major axis of the Earth orbit. The authors of ref \cite{cqg1} choose to fix the {\it eccentricity}  $e$ (the same for all SC's orbits) 

\begin{equation}\label{ecc}
e = \sqrt{1 + \frac43 \alpha^{2} + \frac4{\sqrt{3}} \alpha \cos \theta} - 1
\end{equation}
by  the choice of $\alpha$ and $ \theta$, the inclination of the plane of triangle with respect to the ecliptic,

\be\label{tilt}
\theta=\pm\frac{\pi}3 + \delta=\pm\frac{\pi}3 + \alpha \delta_{1},\ee
where $\delta$ (alternatively $\delta_{1}$) denotes the additional {\it tilt angle} with respect to the canonical value $\pm\frac{\pi}3$. 
Finally, the {\it inclination}  $i$ of the orbital planes with respect to the ecliptic is given by

\begin{equation}\label{inc}
\tan i = \frac2{\sqrt{3}} \frac{\alpha \sin \theta}
                  {1+\frac2{\sqrt{3}} \alpha \cos \theta} .
\end{equation}
We can now write the orbit of one `reference'  spacecraft, e.g. $SC_{1}$, in its parametric form:

\begin{eqnarray}\label{eq:kepler1}
X_{1}&=&R (\cos \psi_{1}+e) \cos i,  \nonumber\\
Y_{1}&=&R \sqrt{1-e^{2}}\sin \psi_{1},  \nonumber\\
Z_{1}&=&R (\cos \psi_{1}+e) \sin i \nonumber,
\end{eqnarray}
The orbits of $SC_{2}$ ed $SC_{3}$, are obtained starting from that of $SC_{1}$ and performing two rotations around the $Z$ axis, of an angle $2 \pi/ 3$ and $4 \pi/3$ respectively, leading to eqs.(\ref{eq:kepler}). The eccentric anomaly as a function of time is given by the solution of the Kepler equation. Actually, we now have {\it three} anomalies, given as solutions of the three equations

\begin{equation}\label{eccentric}
\psi_{k} + e \sin \psi_{k} = \Omega t - \frac{2 \pi}3 (k-1)  \doteq \phi_{k},  \; \; k=1,2,3 
\end{equation}
where the {\it mean motion} is $\Omega = 2 \pi / T = \sqrt{\frac{G M_{\odot}}{R^{3}}}$.
If we measure time in years, then $T=1$ and $ \Omega = 2 \pi $. It is also useful to introduce the phase shifts 

\be\label{sigma}
\sigma_k = \frac{2 \pi}3 (k-1),\ee
so that $\phi_{k} = \Omega t - \sigma_k$. 
The distances among spacecrafts are clearly given by the three lengths

\begin{equation}
L_{jk}=\sqrt{(X_{j}-X_{k})^{2}+(Y_{j}-Y_{k})^{2}+(Z_{j}-Z_{k})^{2}}, \; \; j,k=1,2,3 .
\end{equation}
These distances are not constant in time, causing what is usually referred to as `flexing', even for purely Keplerian orbits. A simple way to see this, is to consider that the instantaneous distance from the Sun are different from one spacecraft to the other. The flexing is quite sensitive to the inclination of the plane of the triangle. The value $\delta_{1}=5/8$, giving the angle $\theta \simeq 60^{\circ} \, 35' \, 45''$
is proved to be optimal by Rajesh Nayak et al.\cite{cqg2}. The maximum flexing of the optimal configuration is about $40000 \; {\rm km}$ and the maximum rate of change of arm lengths is $\pm 4 \; {\rm m/s}$. The reduction of the relative velocity and related Doppler phase shift with respect to the case considered in \cite{cqg2}  ($\delta_{1}=0$) amounts to a factor 5.5. However, it should be noted \cite{benderpc} that flexing figures comparable to the optimal ones can be obtained simply with slightly different choices of the Keplerian orbital parameters \cite{hu,sw}.
The large amount of flexing considered in \cite{cqg2} with $\delta_{1}=0$ is determined by the choice implicit in definitions (\ref{ecc}--\ref{inc}).

\begin{figure}
   \centering
	\includegraphics[width=0.8
	\textwidth]{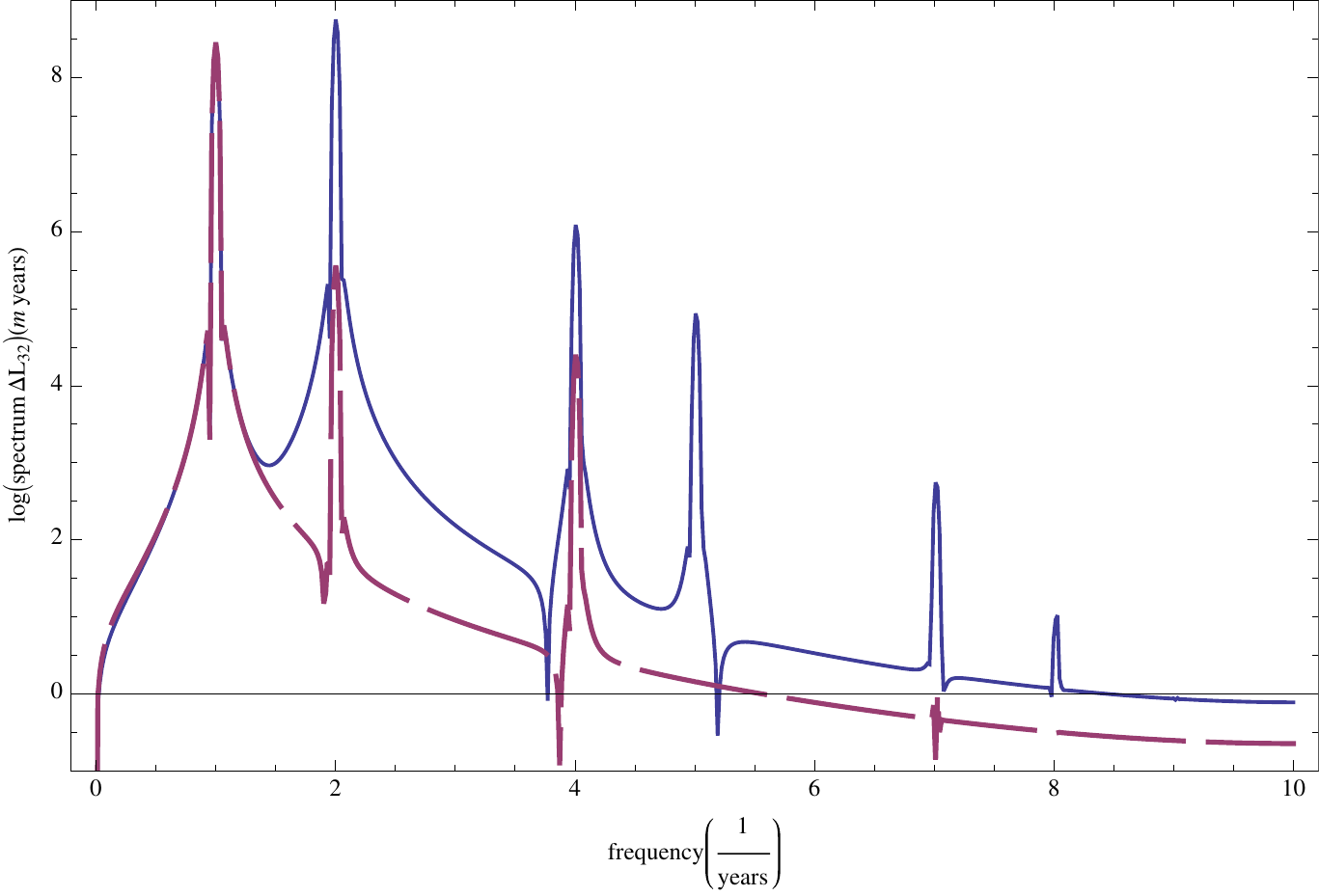}
\caption{Logarithm of the modulus of the Fourier transform of the difference $L_{12}-L_{13}$: standard tilt ($\theta = \pi/3, \;\delta_1 = 0$) (continuous line);  $\delta_{1} = 5/8$ (dashed line).}
         \label{sk}
   \end{figure}

A nice way to appreciate the improvement provided by the optimal choice of the tilt angle is that of comparing the spectral amplitudes of the differential relative motion. Following ref.\cite{cerdo2}, we can compute the spectra of the difference 
 
\be\Delta L_{kj} = L_{ij}-L_{ik}, \quad i,j,k=1,2,3.\ee
In Fig.\ref{sk} the two curves represent the modulus of the Fourier transform of $\Delta L_{32}$ tapered with a Blackman window. The spectrum is computed on a time span of 30 years chosen just to get a good resolution in frequency. The optimal curve has the second harmonic (frequency $2 \, \rm yr^{-1}$) seven orders of magnitudes below that of the standard solution with $ \delta_{1} = 0$. The higher harmonics are much smaller or even unnoticeable. In both solutions the harmonics $3,6,9,...$ are absent in view of the triangular symmetry of the configuration.

\section{The Hill system}

A perturbative analytical approach is convenient when we have to analyze in the simplest way the geometry of orbits and the most relevant perturbations. The Hill reference system (see, e.g. \cite{bp,bvj}, also called as Clohessy-Wiltshire system \cite{nerem}) allows us to reduce the equations of motion to those of a set of coupled oscillators. The Hill equations properly said consist of the linear part of this system which is readily solved. Non-linear and non-autonomous terms can be added and solved perturbatively. In the specific instance, the system is at rest with the barycenter of the constellation and so an observer sees the Earth at rest in it. The new reference is introduced according to the following steps:

1. Start by considering a circular orbit of radius $R$ around the Sun and pass from the inertial barycentric system to a new system with its origin rotating with the reference orbit and with axes $x=x_{1},y=x_{2},z=x_{3}$. $x_{1}$ is directed radially opposed to the center, $x_{2}$ is in the direction tangent to the motion and $x_{3}$ is perpendicular to their plane. The system is non-inertial and the equations of motion are

\begin{equation}\label{hill}
\ddot {\bv x} = - \nabla \Phi - 2 \bv \Omega \wedge \dot {\bv x}
- {\bv \Omega} \wedge \left({\bv \Omega} \wedge ({\bv x} + \bv R) \right).
\end{equation}
$\Phi$ is the total gravitational potential and ${\bv \Omega} = (0,0,\Omega), {\bv R} = (R,0,0)$. 

2. The potential is approximated according to the expansion

\begin{equation}\label{hill2}
\Phi = \Phi \vert_{\bv x=0} + 
   \frac{\partial \Phi}{\partial x_{\lambda}} \bigg\vert_{\bv x=0} x_{\lambda} 
+ \frac12 \frac{\partial^{2} \Phi}{\partial x_{\mu} \partial x_{\lambda}} \bigg\vert_{\bv x=0} x_{\mu} x_{\lambda} 
+ {\rm O} (|\bv x|^{3}),
\end{equation}
where the sum is implicit over the repeated indexes $\lambda, \mu =1,2,3$. The ``far tide'' (or Hill) approximation \cite{bp} consists of retaining only the first order term into the expansion of the gradient of the potential. Taking

\begin{equation}
\Phi=-\frac{G M_{\odot}}{\sqrt{(x+R)^{2}+y^{2}+z^{2}}}
\end{equation}
the {\it Solar} far tide approximation gives

\begin{equation}
\frac{\partial\Phi}{\partial x} \bigg\vert_{\bv x=0}=\frac{G M_{\odot}}{R^{2}}=\Omega^{2} R, \;\;\;\;
\frac{\partial\Phi}{\partial y} \bigg\vert_{\bv x=0} = \frac{\partial\Phi}{\partial z} \bigg\vert_{\bv x=0} = 0 \end{equation}
and

\begin{equation}
\frac{\partial^{2} \Phi}{\partial x^{2}} \bigg\vert_{\bv x=0}=-
\frac{2 G M_{\odot}}{R^{3}}=-2 \Omega^{2}, \;\;\;\;
\frac{\partial^{2} \Phi}{\partial y^{2}} \bigg\vert_{\bv x=0} = 
\frac{\partial^{2} \Phi}{\partial z^{2}} \bigg\vert_{\bv x=0} = \frac{G M_{\odot}}{R^{3}}= \Omega^{2}, \end{equation}
all others zero. The next order of approximation ({\it octupole}) is:

\begin{eqnarray}
\frac{\partial^{3} \Phi}{\partial x^{3}} \bigg\vert_{\bv x=0}&=&
\frac{6 G M_{\odot}}{R^{4}}=\frac{6\Omega^{2}}{R}, \\
\frac{\partial^{3} \Phi}{\partial x \partial y^{2}} \bigg\vert_{\bv x=0}&=&
\frac{\partial^{3} \Phi}{\partial x \partial z^{2}} \bigg\vert_{\bv x=0}=-
\frac{3 G M_{\odot}}{R^{4}}=-\frac{3\Omega^{2}}{R}, 
\end{eqnarray}
all others zero. By using these in (\ref{hill}) we get

\begin{eqnarray}
\ddot x &-& 2 \Omega \dot y - 3 \Omega^{2} x + \frac{3\Omega^{2}}{2R} (2 x^{2} - y^{2} - z^{2}) =0,\nonumber \\
\ddot y &+& 2 \Omega \dot x - \frac{3\Omega^{2}}{R} xy =0,\\
\ddot z &+& \Omega^{2} z - \frac{3\Omega^{2}}{R} xz=0 \nonumber.
\end{eqnarray}
These are the equations for a second order approximation of the purely Keplerian motion. Any other perturbing force can be derived by an analogous series expansion of the corresponding potential and added to this system. Each component, coherently with a perturbative approach, will be a power series in the coordinates with suitably `small' coefficients ${\bv \varepsilon} = (\varepsilon_x, \varepsilon_y, \varepsilon_z)$. Recalling the definition of the parameter $\alpha$ in (\ref{alfa}) and introducing explicitly only additional {\it autonomous} perturbations represented by the perturbation vector $\bv f (\bv x, \bv \varepsilon)$, the system (\ref{hill}) at second order can be written in  the form

\begin{eqnarray}\label{eq2_123}
\ddot x &-& 2 \Omega \dot y - 3 \Omega^{2} x + 
\frac{3 \Omega^{2} \alpha}{\ell} (2 x^{2} - y^{2} - z^{2}) - f_x (\bv x, \bv \varepsilon)=0,\nonumber \\
\ddot y &+& 2 \Omega \dot x - \frac{6 \Omega^{2} \alpha}{\ell} x y - f_y (\bv x, \bv \varepsilon)=0,\\
\ddot z &+& \Omega^{2} z - \frac{6 \Omega^{2}\alpha}{\ell} x z - f_z (\bv x, \bv \varepsilon)=0. \nonumber
\end{eqnarray}
Since $\alpha$ is small, the nonlinear terms in the multipole expansion of the Solar field can be included in the perturbations. This is an important remark, since in general the presence of quadratic or higher-order terms in the system  (\ref{eq2_123}) makes it {\it non-integrable}. The most effective approach is that of finding the exact solution of the linear part and using it recursively to find approximate solution for the full non-linear system.

The perturbation vector $\bv f (\bv x, \bv \varepsilon)$ could even start simply with linear terms of the form $ \varepsilon x + \dots $ giving explicit small changes to the unperturbed zero-order solution. However, it is more natural to consider them of magnitude comparable to terms of order $\alpha^2$ or higher. These and possibly higher-order perturbing terms propagate in the full solutions through the non-linear coupling with the other components of the field expansion. The solution of the system can therefore be written in the form of a series expansion

\begin{eqnarray}\label{approxxyz}
x_{k}(t) = x_{k}^{(0)} + \alpha x_{k}^{(1)} + \varepsilon_x x_{k}^{(2)} + \dots, \nonumber \\
y_{k}(t) = y_{k}^{(0)} + \alpha y_{k}^{(1)} + \varepsilon_y y_{k}^{(2)} + \dots, \\
z_{k}(t) = z_{k}^{(0)} + \alpha z_{k}^{(1)} + \varepsilon_z z_{k}^{(2)} + \dots, \nonumber 
\end{eqnarray}
where the value of the upper index denotes the order and with the index $k=1,2,3$ we denote as usual the three spacecrafts. 

The `zero' order solution is obtained by putting $\alpha= \bv \varepsilon =0$ in (\ref{eq2_123}). It amounts to be

\ba
x_1(t) &=& 2 \left( 2 x_1 (0)  + \frac{1}{\Omega} \dot y_1 (0) \right) + \frac{1}{\Omega} \dot x_1 (0) \sin \Omega t - 
                     \left( 3 x_1 (0)  + \frac{2}{\Omega} \dot y_1 (0) \right) \cos \Omega t,  \nonumber \\
y_1(t)  &=& y_1 (0)  - \frac{2}{\Omega} \dot x_1 (0)  \\
&& - 3 \left( 2 \Omega x_1 (0)  + \dot y_1 (0) \right) t + \frac{2}{\Omega} \dot x_1 (0) \cos \Omega t + 2  
                     \left( 3 x_1 (0)  + \frac{2}{\Omega} \dot y_1 (0) \right) \sin \Omega t,  \nonumber \\
z_1(t)  &=& z_1 (0) \cos \Omega t + \frac{\dot z_1 (0)}{\Omega} \sin \Omega t. \nonumber
\ea
The solutions for the other two spacecrafts are obtained by substituting $\Omega t$ with $\phi_{k}$ as in definition (\ref{eccentric}).
However, the solution actually employed in the perturbation expansion (the {\it epicyclic} motion) is obtained by choosing initial conditions in order to remove constant and secular terms \cite{cqg1}:

\begin{equation}\label{zeroxyz}
x_{k}^{(0)} = \frac{\ell}{2\sqrt{3}} \cos \phi_{k}, \hskip 1.5cm
y_{k}^{(0)} = -\frac{\ell}{\sqrt{3}} \sin \phi_{k}, \hskip 1.5cm
z_{k}^{(0)} = \frac{\ell}{2} \cos \phi_{k}. 
\end{equation}
It corresponds to harmonic oscillations around the `reference' $\bv x =0$ circular orbit. 

The order `one' (in $\alpha$) solution with the integration constants chosen in order to comply with the LISA constraints is \cite{cqg2}

\begin{eqnarray}
x_{k}^{(1)} &=& -\frac{5\ell}{24} + \frac{\ell}2 \left(\frac12 - \delta_{1}\right) \cos \phi_{k} - 
                          \frac{\ell}{24} \cos 2\phi_{k}, \nonumber \\
y_{k}^{(1)} &=& -\ell \left(\frac12 - \delta_{1}\right) \sin \phi_{k} + 
                          \frac{\ell}{6} \sin 2\phi_{k},  \\
z_{k}^{(1)} &=& \frac{\sqrt{3}\ell}{4} - \frac{\ell}{2\sqrt{3}} \left(1 - \delta_{1}\right) \cos \phi_{k} - 
                          \frac{\ell}{4\sqrt{3}} \cos 2\phi_{k}. \nonumber
\end{eqnarray}
The solution 

\be\label{approx}
{\bv x}_{k}(t) = {\bv x}_{k}^{(0)} + \alpha {\bv x}_{k}^{(1)}\ee 
with these explicit forms is a very good approximation of the exact Keplerian orbits, with an error less than $0.03 \%$ on the prediction of the flexing of the arms \cite{cqg2,sw}. In other terms, the plots of Fig.\ref{sk} can be reproduced using solution (\ref{approx}) with a remarkable accuracy. This result proves that this solution is a reliable starting point for the study of additional sources of gravitational disturbances. From the analytical point of view it is worth mentioning that also system (\ref{eq2_123}) with just $\bv f (\bv x, \bv \varepsilon) \equiv {\bv 0}$ (of which (\ref{approx}) is a first order approximate solution) is not integrable unlike the original Kepler system which is clearly integrable. Is a common lesson of modern analytical mechanics that truncated solutions of non-integrable approximations can be more useful than formally exact solutions of their integrable counterparts.

\section{The Earth effect}

As mentioned above the Earth is practically at rest in the Hill system. The Earth coordinates are

\begin{equation*}
x_{\oplus}=-R(1-\cos 20^{\circ})\simeq-9.0\times 10^6 \; {\rm km}, \;\;
y_{\oplus}=R\sin 20^{\circ}\simeq5.13\times 10^7 \; {\rm km}, \;\; z_{\oplus}=0.
\end{equation*}
We can therefore consider its effect as a {\it constant} + {\it linear} perturbation to add to the equations of motion. If
$d_{\oplus}=\sqrt{x_{\oplus}^{2}+y_{\oplus}^{2}}\simeq 5.2\times 10^7 \; {\rm km},$
is the distance from the origin, we can add the perturbative parameter

\be
\varepsilon_{\oplus} = \frac{M_{\oplus}}{M_{\odot}} \left(\frac{R}{d_{\oplus}}\right)^{3}\simeq 7.3\times 10^{-5},\ee
where we used the ratio 
\begin{equation*}
\frac{M_{\oplus}}{M_{\odot}} = \frac1{328900}.
\end{equation*}
The equations of motion can now be written in the form (\ref{eq2_123}) with the perturbation vector $\bv f (\bv x, \bv \varepsilon)$ given by

   \begin{align}
\bv f_{\oplus}  &= - \varepsilon_{\oplus} \Omega^{2} ({\bv x}-{\bv x}_{\oplus}).\label{eqt1}
\end{align}
Proceeding as assumed in eqs.(\ref{approxxyz}), we may look for a solution that can be linearly superposed to (\ref{approx}) so that we have

\be\label{approx2}
{\bv x}_{k}(t) = {\bv x}_{k}^{(0)} + \alpha {\bv x}_{k}^{(1)} + \varepsilon_{\oplus} {\bv x}_{k}^{(2)}. \ee
The explicit solution is \cite{cqg3}:

\ba\label{seqt}
x_{k}^{(2)} &=& 2  \frac{A_{k}}{\Omega} + x_{\oplus} + 2 y_{\oplus} \phi_{k} + 
 B_{k} \cos \phi_{k} + C_{k} \sin \phi_{k} - \frac{5\ell}{4\sqrt{3}} \phi_{k} \sin \phi_{k}, \nonumber \\
y_{k}^{(2)} &=& -(3 \frac{A_{k}}{\Omega} + 2 x_{\oplus}) \phi_{k} - \frac32 y_{\oplus} \phi_{k}^{2} + \\
&& -\frac{5\ell}{2\sqrt{3}} \phi_{k} \cos \phi_{k} +
                         \frac{\sqrt{3} \ell}2 \sin \phi_{k} + 2 (C_{k} \cos \phi_{k} - B_{k} \sin \phi_{k}) + D_{k}, \nonumber \\
z_{k}^{(2)} &=& E_{k} \cos \phi_{k} + F_{k} \sin \phi_{k} - \frac{\ell}4 \phi_{k} \sin \phi_{k}, \nonumber
\ea
where the $A_k,...,F_k$ are constants determining the initial conditions. Recalling that the mean anomalies $\phi_{k}$ are just phase shifted rescaling of time, in these solutions we see the presence of additional secular terms. In particular, the $y$ coordinate has a term quadratic in time that cannot be eliminated by any choice of the initial conditions.

\begin{figure}
   \centering
\includegraphics[width=0.9\textwidth]{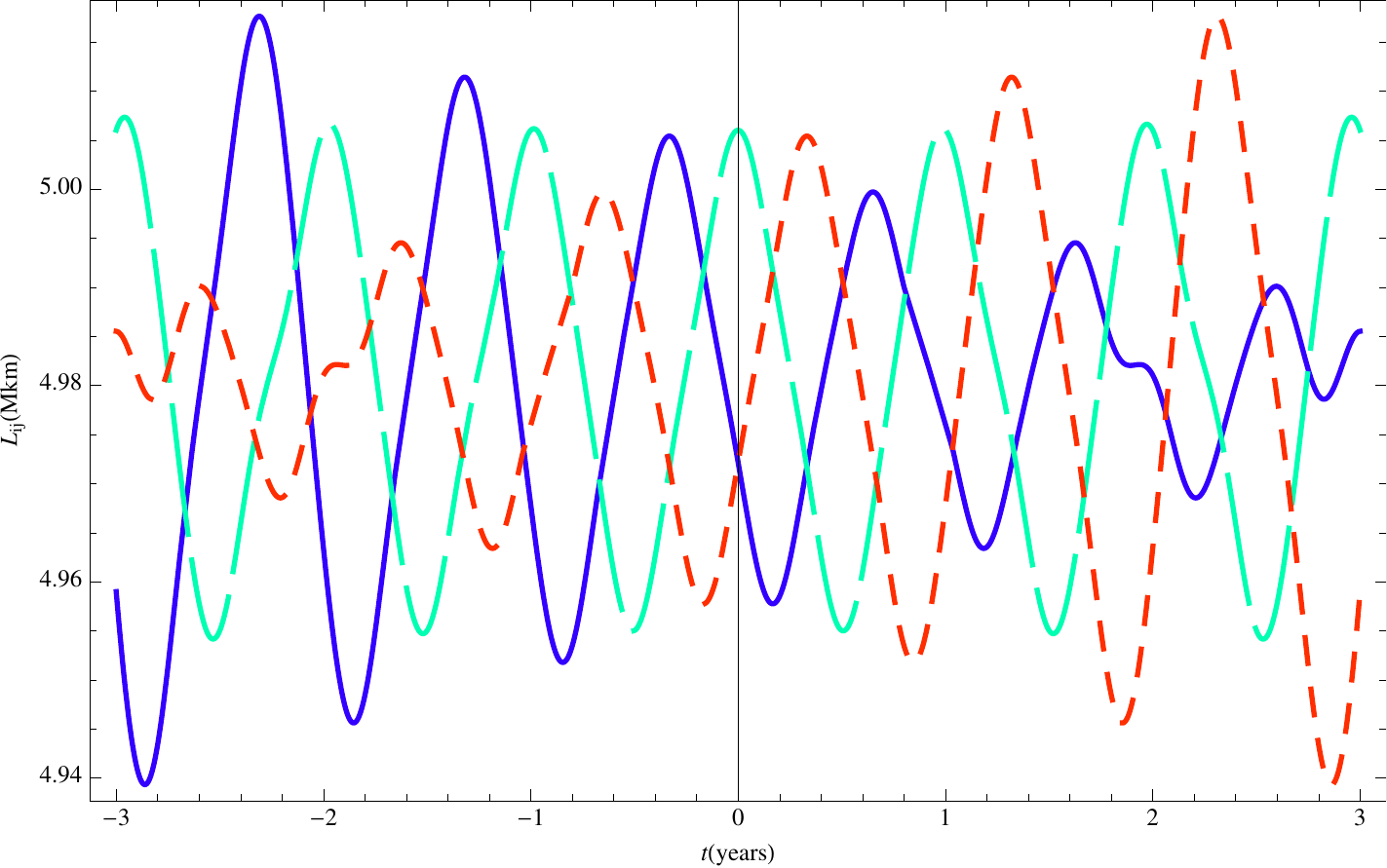}
      \caption{Distances $L_{ij}$ vs time including the Earth influence for a period of 6 years: $L_{12}$, continuous line; $L_{23}$, long dashes; $L_{13}$ short dashes.}
         \label{terra1}
     \end{figure}   

  \begin{figure}
   \centering
\includegraphics[width=0.9\textwidth]{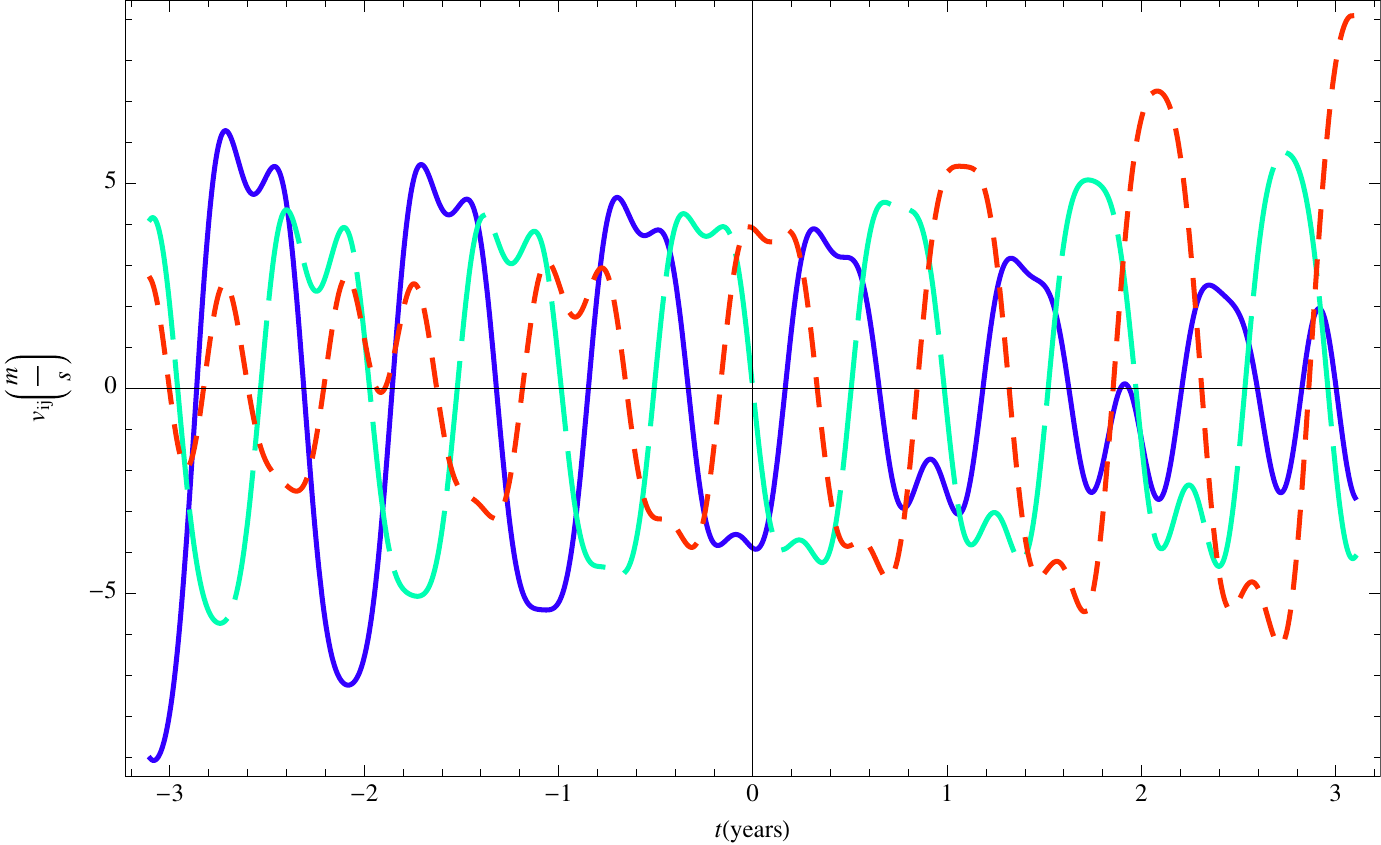}
      \caption{Rates of change $v_{ij}=\dot L_{ij}$ vs time including the Earth influence for a period of 6 years: $v_{12}$, continuous line; $v_{23}$, long dashes; $v_{13}$ short dashes.}         \label{terra2}
   \end{figure}   

Dhurandhar et al.\cite{cqg3} propose to inject LISA on orbits with initial conditions devised to reduce to a minimum the Earth effect. In view of the secular terms, it seems a good strategy that of minimizing the disturbance at `mid mission': they assume a mission duration of three years, so that this `optimal time' is at 1.5 years after starting the operation activity. The authors of ref.\cite{cqg3} introduce in the mean anomalies $\phi_{k}$ a phase shift to accommodate for different starting times and compute the resulting effect on the chosen period. However, the solution is invariant under time translation in view of the autonomous nature of the system, therefore, expressions (\ref{seqt}) can be profitably used at arbitrary initial time. If this, for simplicity, is put equal to zero, the values of the constants appearing in  (\ref{seqt}) are

\ba\label{init6}
A_{k} &=& \Omega y_{\oplus} \sigma_{k} + \frac{\Omega \ell}{\sqrt{3}} \cos \sigma_{k} ,  \nonumber \\
B_{k} &=& -\frac{13 \ell}{8 \sqrt{3}} - x_{\oplus} \cos \sigma_{k} - 2 y_{\oplus} \sin \sigma_{k} - \frac{\sqrt{3} \ell}{8} \cos 2 \sigma_{k}, \nonumber \\
C_{k} &=& x_{\oplus} \sin \sigma_{k} - 2 y_{\oplus} \cos \sigma_{k} + \ell \frac{3 \sin 2 \sigma_{k} - 10 \sigma_k}{8 \sqrt{3}}, \\
D_{k} &=& y_{\oplus} \left(4 - \frac32 \sigma_{k}^2 \right) - 2 x_{\oplus} \sigma_{k}  - \sqrt{3} \ell \sigma_{k} \cos \sigma_{k} + \frac{4 \ell}{\sqrt{3}} \sin \sigma_{k} , \nonumber \\
E_{k} &=& -\frac{\ell}{4} \sin^2 \sigma_{k} , \hskip 2cm%  \nonumber \\
F_{k} = -\frac{\ell}{8} \left( 2 \sigma_{k}  - \sin 2 \sigma_{k} \right), \nonumber 
\ea
where the $\sigma_{k} $ are the phase shifts introduced in (\ref{sigma}). 

Experience gathered on many space experiments of the past allows some optimism on expecting the LISA mission to last longer than the nominal duration of three years.  A longer lifetime also allows us to better investigate the secular (non periodic) behavior of the orbits. In order to clarify the effect of the Earth tidal field, 
we have therefore chosen to consider an ``optimistic duration" and to double the expected mission time: 
in Fig.\ref{terra1} the three arm lengths  time are plotted over a period of 6 years with the solutions (\ref{approx2}) and the initial conditions above. In Fig.\ref{terra2} they are plotted, on the same six years period, the relative velocities $v_{jk}=\dot L_{jk}$. Now we can see that, if we limit the observation to the three years centered around zero, both the flexing and the relative velocity have an almost constant envelope. However, in the long run the envelope tends to linearly grow for both quantities and is therefore appropriate a comparison with the `optimal' solution illustrated in section 2. In a three-year ($\pm 1.5$) mission we have a flexing of $60000 \; {\rm km}$ which increases to $80000 \; {\rm km}$ in six  ($\pm 3$) years: on this longer time span we arrive at a factor 2 of increase of the flexing with respect to the purely Keplerian solution. The maximum rate of change of arm lengths is $\pm 5.5 \; {\rm m/s}$ in the $\pm 1.5$ years time span but increases to $\pm 10 \; {\rm m/s}$ in the $\pm 3$ years time span, a factor 2.5 of increase with respect to the Keplerian solution. 

\begin{figure}
   \centering
\includegraphics[width=0.8\textwidth]{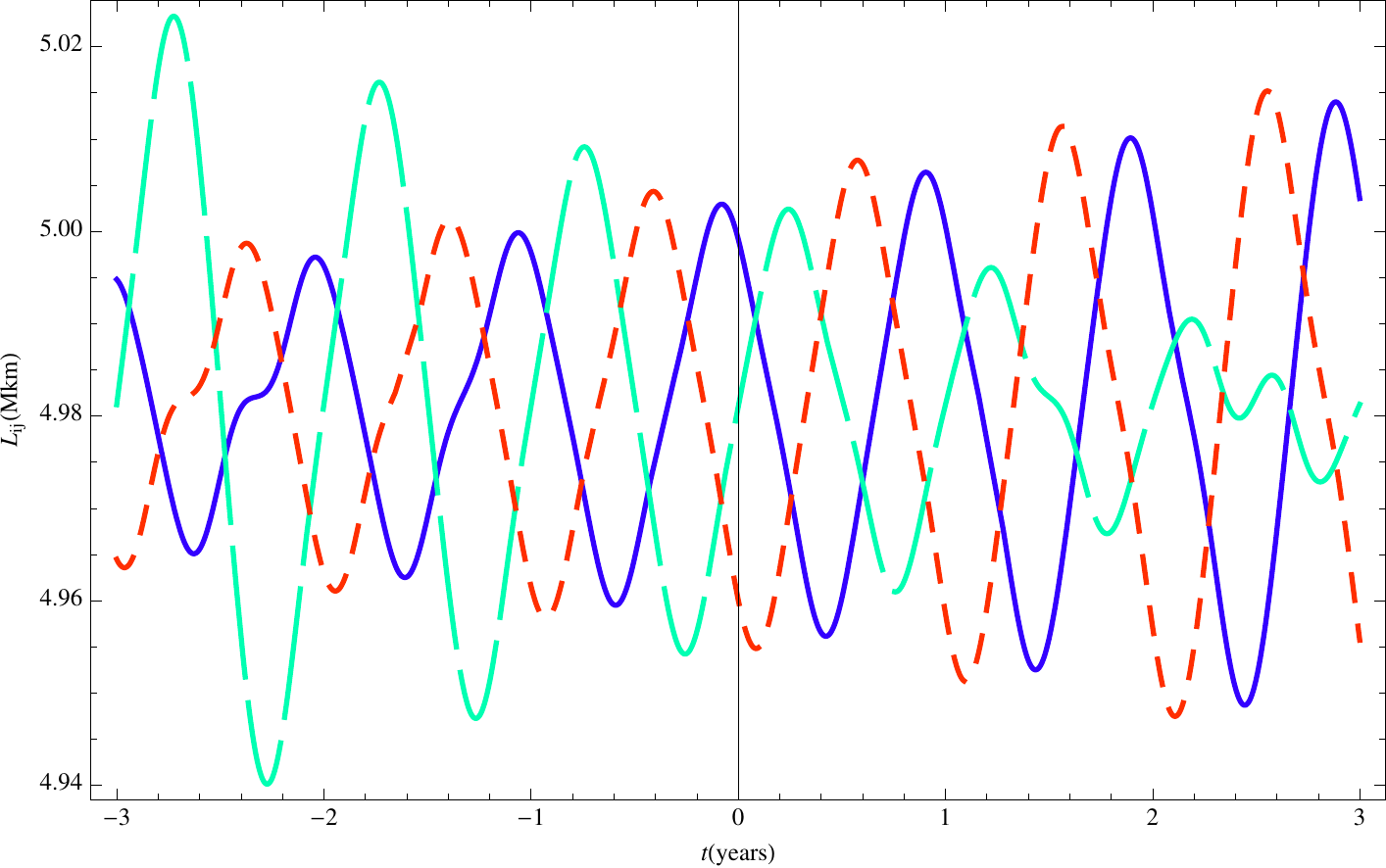}
      \caption{Distances $L_{ij}$  time including the Earth influence for a period of 6 years with an extra $\gamma=-\pi/2$ phase: $L_{12}$, continuous line; $L_{23}$, long dashes; $L_{13}$ short dashes.}         \label{terra6}\end{figure}   
 
This behavior is due to the global dynamics in the coupled fields of Sun and Earth as described in the {\it rotating frame}. As can be seen from the solution (\ref{seqt}) with the constants determined by (\ref{init6}), the secular terms always produce a drift {\it away} from Earth: this is linear in $x$ and, as remarked above, quadratic in $y$. The secular drift in $x$ of the three $SC_k$ amounts to $\pm 70000 \; {\rm km}$ on the $\pm 1.5$ years time span, whereas the motion in $y$ is of initial approach, a state of rest and a subsequent departure: on the same time span, the drift in $y$ is of $5 \times 10^5 \; {\rm km}$. So, the best that we can do is a choice of initial conditions such that LISA is a little further out at start, approaches the Earth, is at rest ad mid mission and departs after that. The physical reason of this behavior is evident when we consider the nature of the effective potential. LISA is well outside the Roche lobe associated to the Earth; therefore it cannot `fall' (as a stone or a projectile \cite{cqg3}) towards it. Rather, it moves on a slow merry-go-round and feels a small Coriolis apparent force. The combined parabolic-drift motions have therefore quite a relevant role in increasing the flexing on a certain time base. 

However, the possibility exists of an appreciable reduction of the flexing due to the Earth tidal field over a limited time span by a suitable shift in the relative phase between the epicyclic motion and the linear perturbation. The scenario depicted above can in fact be changed if the constellation gets to the point of minimum distance from the Earth with a different orientation. This situation can be described either by modifying the phase shift in the initial conditions of the linear perturbation or by putting this extra shift directly in the definition (\ref{sigma}). Let us modify this in 

\be\label{ss}
\hat\sigma_k = \frac{2 \pi}3 (k-1)-\gamma_k, \; \; k=1,2,3 \ee
using the angles $\gamma_k$ to change the orientation of the triangle at a given time. We limit the analysis to identical angles for all the $SC_k$, however a non-rigid rotation could also be studied. The standard zero order solution used above, with $\gamma=0$, of (\ref{zeroxyz}) corresponds to a configuration with the spacecraft 1 at the `highest' point ($z_{1} = \ell / 2$) over the ecliptic at time zero (let us say at mid mission). The spacecraft 1 being at $z_{1} = 0$ with either $y_{1} = + \ell / \sqrt{3}$ or $y_{1} = - \ell / \sqrt{3}$ correspond to the two extremes in which it is closest or farthest from the Earth at time zero. It is interesting to examine these two situations compared with that above: they are given by putting $\gamma=\pm\pi/2$ in (\ref{ss}). In Fig.\ref{terra6} the first case ($\gamma=-\pi/2 $) is illustrated by plotting the three arm lengths  time over the period of 6 years: in this case in which $SC_1$ is the one closest to the Earth at time zero, the two arms 12 and 13 exhibit a variation not larger than the optimal one ($40000 \; {\rm km}$) for the first half-mission period of 3 years. It can be shown that in the second case ($SC_1$ farthest from  the Earth at time zero) the plot is the time reversal of that in Fig.\ref{terra6}, so that the variation of $L_{12},L_{13}$ is kept to a minimum in the second 3 years. The price for this reduction is that the flexing of the third (23) arm is higher, but only slightly with respect to the level found with $\gamma=0$ ($80000 \; {\rm km}$) on the same time span. 

\begin{figure}
   \centering
\includegraphics[width=0.8\textwidth]{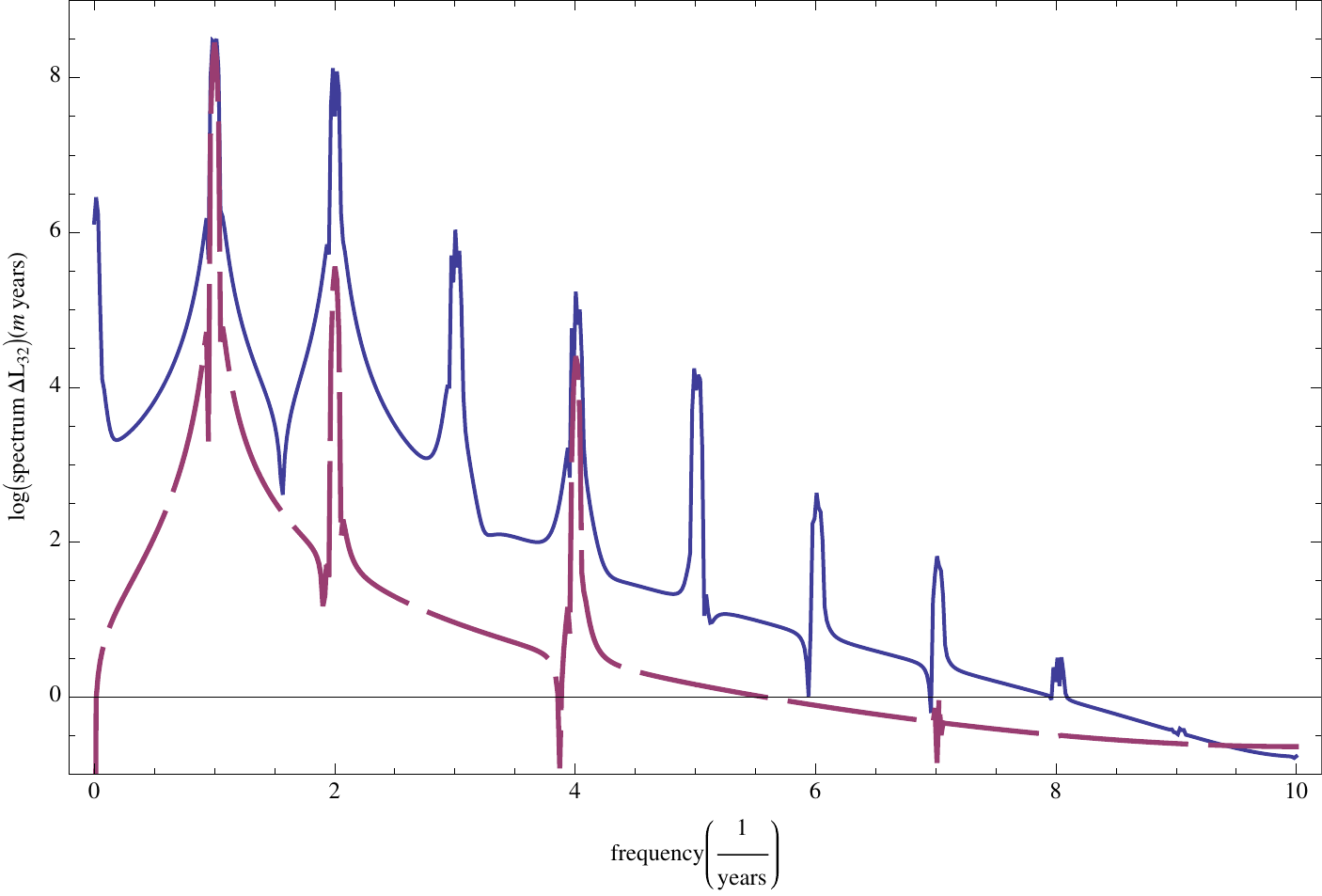}
      \caption{Spectrum of the difference $\Delta L_{32} = L_{12}-L_{13}$ for the case $\gamma=0$. The dashed line is the Keplerian configuration and the continuous line represents the addition of the Earth perturbation.}
         \label{stz}\end{figure}

Another way to examine the effect of the Earth is by comparing the differential relative motion computing the difference $\Delta L_{32} = L_{12}-L_{13}$ and the corresponding spectral amplitudes. In Fig.\ref{stz} the modulus of the Fourier transform of $\Delta L_{32}$ is shown together with the optimal Keplerian case of Fig.\ref{sk} (no Earth contribution). The continuous line includes the contribution of the Earth in the case of $\gamma=0$. As before, to get a good resolution in frequency, the spectrum is computed on a longer time span (30 years). The effect of the Earth is evident from the appearance of the harmonics $3,6,9,...$ due to the fact that the triangular symmetry of the configuration is now broken. The configurations with the phases $\gamma=\pm \pi/2$ display the same spectral lines with slightly lower values at higher frequencies.

The analysis of these and other possible configurations could be of a certain practical relevance should it be necessary to privilege the proper working of only a pair of arms \cite{drv10}.

\section{The relativistic precession and the effect of the Solar quadrupole}

In the list of small perturbative effects given by fixed fields naturally fall those connected with the not exactly Keplerian interaction with the Sun. Two important examples are the relativistic correction of the classical field and the perturbation due to a possible quadrupole moment of the Solar gravitational field. The first produces the slow precession responsible for the advance of the perihelion; the second produces a more complex variation of orbital elements but of (supposedly) much smaller amount.

It is tempting to see if LISA is capable to detect small effects due to the violation of the spherical symmetry of the Solar field, such as we have assumed so far. The small eccentricity and the distance from the Sun of Earth-like orbits makes the amount of the effects quite small, but worth investigating in view of the high sensitivity of LISA to small differential variations of the arm lengths. 
On this ground there is the recent proposal by Polnarev et al.\cite{PRB} of detecting with LISA the low frequency g-mode Solar oscillations.
 
The relativistic correction can be mimicked by the presence of a small additional term in the potential of the form

\be\label{Schw}
\Phi_{\rm S} = - \frac{r_{\rm S} L^2}{r^3}, \nonumber \ee
where
 $$r_{\rm S} = \frac{GM_{\odot}}{c^2}$$ 
 is the {\it Schwarzschild radius} of the Sun and $L$ is the angular momentum of the orbit related to the other parameters by $L^2 = GM_{\odot} R (1-e^2)$. The Solar quadrupole is associated with a small term of the form

\be\label{Solar}
\Phi_{\rm Q} = \frac{GM_{\odot} J_2 R_{\odot}^2}{r^3}\cdot \frac{2 Z^2 - X^2 - Y^2}{2 r^2},\ee
where $J_2 =  Q \times10^{-7}$
characterizes the mass quadrupole of the Sun, $Q$ being a parameter of order $1$ according to current estimates \cite{HR}. $R_{\odot} \simeq 7 \times 10^5$ km is the radius of the Sun and the radial distance is expressed in terms of the coordinates as 

$$r = \sqrt{X^{2}+Y^{2}+Z^{2}} = \sqrt{(R+x)^{2}+y^{2}+z^{2}}.$$

An approximate evaluation of the two effects can be attempted in the Hill frame with the usual hypothesis of small perturbations of almost circular orbits. The potentials with the perturbing terms introduced above are expanded in series about the origin of the Hill frame and the first and second order terms are retained and added to the Keplerian terms. We remark that a small constant term appears in the $x$ component because the centrifugal force in the rotating frame exactly balances only the Keplerian monopole at $r=R$: the extra term is responsible for the classical precession effect common to all spacecrafts. The linear terms are given by the different quadrupole terms arising in the expansions: these give significant contributions in the form of differential effects. The equations of motion are of the same form for both effects and can be written in the form (\ref{eq2_123}) with the perturbation vector $\bv f (\bv x, \bv \varepsilon)$ given by

\be f_x = - \varepsilon_x \Omega^{2} \left( R - 4x \right),\quad 
     f_y = - \varepsilon_y \Omega^{2} y,\quad
     f_z = -  \varepsilon_z \Omega^{2} z.
\ee
The Schwarzschild correction is obtained if the coupling constants are given by

\be\label{Schc}
\varepsilon_x = \varepsilon_y = \varepsilon_z = \varepsilon_{\rm S} \equiv 3 \frac{r_{\rm S}}{R} \simeq 3 \times 10^{-8}\ee
whereas the Solar quadrupole is included when the coupling constants are given by

\be\label{Solc}
\varepsilon_x = \varepsilon_y = \frac13 \varepsilon_z = \varepsilon_{\rm Q} \equiv \frac32 J_2 
\left(\frac{R_{\odot}}{R}\right)^2  \simeq 3.3 \ Q \times 10^{-12}.\ee 

\begin{figure}
   \centering
\includegraphics[width=0.8\textwidth]{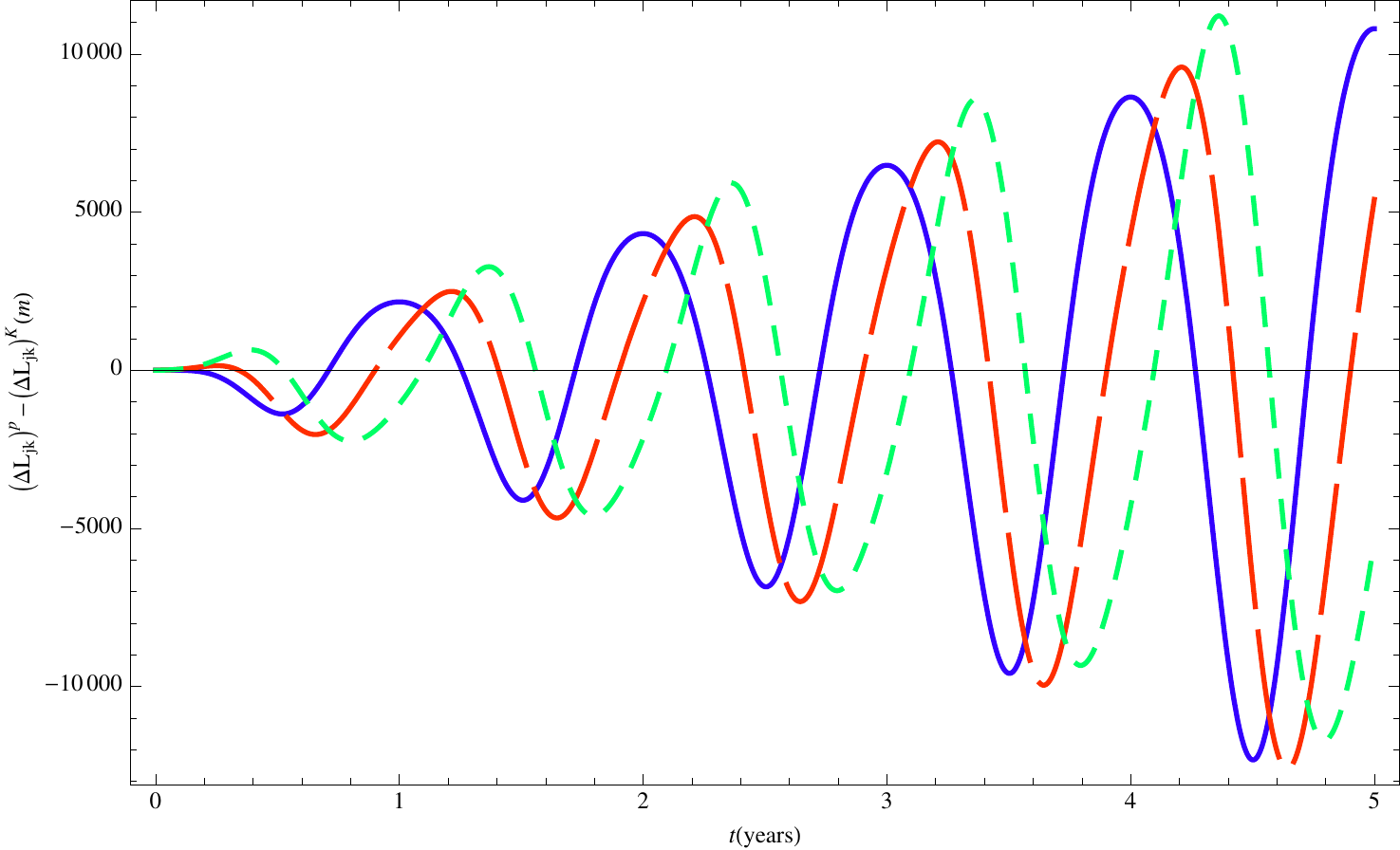}
      \caption{Difference $\Delta L_{jk}^{pert}-\Delta L_{jk}^{kepl}$  time for a period of 5 years as a consequence of the differential effect due to the relativistic precession: the continuous line is for $jk = 23$, the long dashed line for $jk = 13$ and the short dashed  line for $jk = 12$.}
         \label{pr}\end{figure}
         \begin{figure}
   \centering
\includegraphics[width=0.8\textwidth]{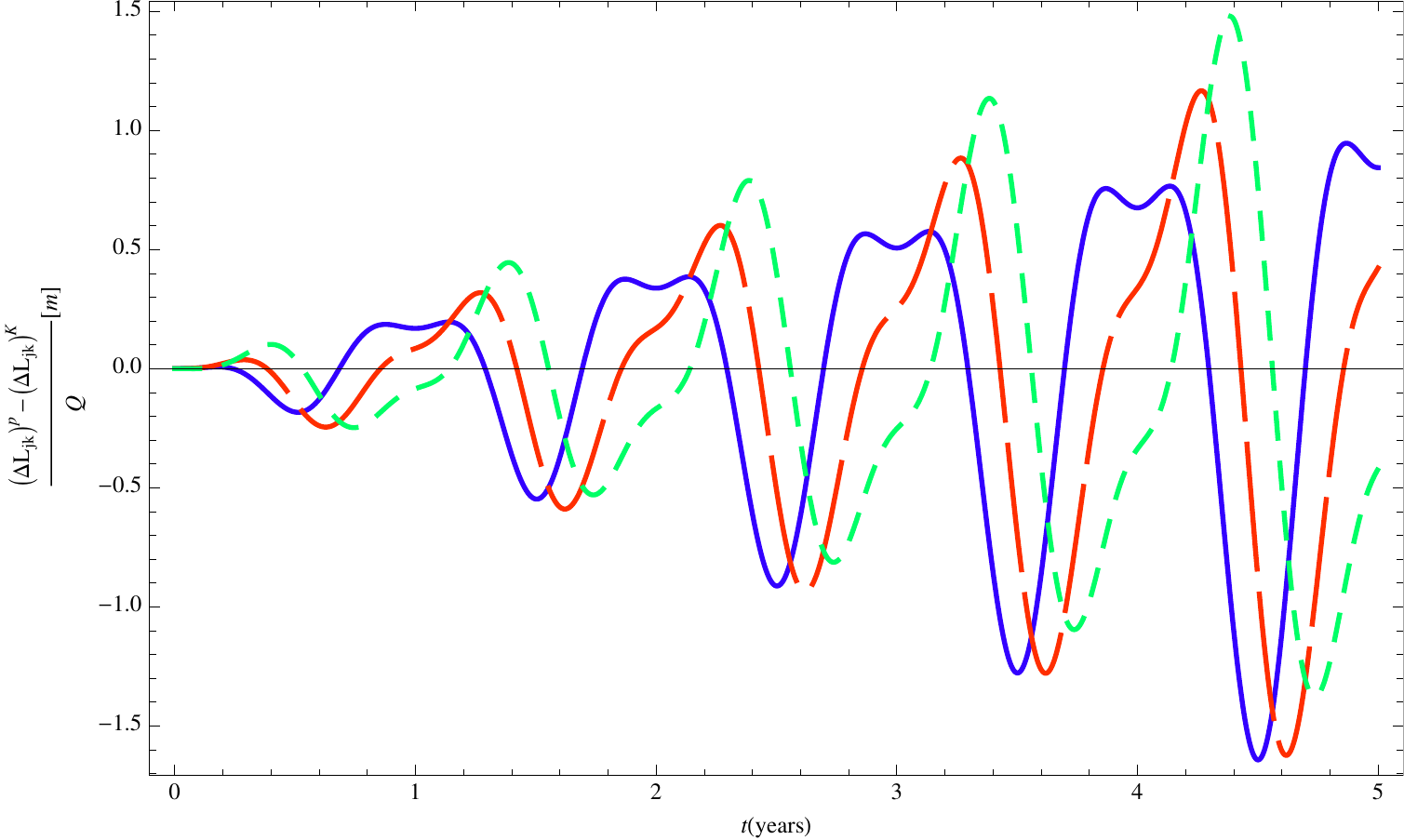}
      \caption{Difference $\Delta L_{32}^{pert}-\Delta L_{32}^{kepl}$ in units of $Q$  time for a period of 5 years as a consequence of the differential effect due to the Solar quadrupole: the continuous line is for $jk = 23$, the long dashed line for $jk = 13$ and the short dashed line for $jk = 12$.}
         \label{pq}\end{figure}

The structure of the system is the same as in the case of the Earth (cfr. Eqs.(\ref{eq2_123}) with the perturbation vector given by (\ref{eqt1})) and the strength of the perturbing terms can be evaluated by comparing the above constants to $\varepsilon_{\oplus} \simeq 7.3\times 10^{-5}$. For the consistency of the approach we preliminarily check the prediction concerning the overall precession.
The constant force appearing in the $x$ component is a factor $R/\ell = (2 \alpha)^{-1} = 30$ stronger than that linearly depending on the coordinates. Therefore we can assign to the solution the following structure

\be\label{approx3}
{\bv x}_{k}(t) = {\bv x}_{k}^{(0)} + \alpha {\bv x}_{k}^{(1)} + \frac{\varepsilon_x}{2\alpha} {\bv x}^{(2)} + {\bv x}_{k}^{(3)} (\bv \varepsilon). \ee
In the second-order term there is no index $k$ because the solution satisfying proper initial conditions on the epicycle is the same for each spacecraft. The structure of this term has to comply with the natural condition that the radial oscillation be symmetric with respect to the circular reference orbit whose radius is slightly reduced by the constant component of the force. The choice of  initial conditions 

\ba
x_k (0) &=& \frac{\ell}{2\sqrt{3}}  \cos \sigma_{k}, \;\;\; 
y_k (0) = \frac{\ell}{\sqrt{3}}  \sin \sigma_{k}, \;\; \;
z_k (0) = \frac{\ell}{2}  \cos \sigma_{k}; \nonumber\\
\dot x_k (0) &=& \frac{\Omega \ell}{2\sqrt{3}}  \sin \sigma_{k}, \;\; \;
\dot y_k (0) = -\frac{\Omega \ell}{\sqrt{3}} \left(   \cos \sigma_{k} - \frac{\varepsilon_x}{2\sqrt{3}\alpha} \right), \;\;\; 
\dot z_k (0) = \frac{\Omega \ell}{2}  \sin \sigma_{k}, \nonumber
\ea
gives the correct solution

\be\label{pre2}
x^{(2)} (t) = -\frac{\ell}{3}\left(1-\cos\Omega t \right),\quad
y^{(2)} (t) = \ell \left(\Omega t-\frac23 \sin\Omega t \right), \quad
z^{(2)} (t) =0,
\ee
since the constant part in $x(t)$, $
-\varepsilon_x \ell (6\alpha)^{-1} = -\varepsilon_x R/3$,
accounts for the change of radius of the reference orbit. An estimate of the advance of the perihelion after a revolution is given by

\be
\Delta \varphi \simeq \frac{\varepsilon_y}{2\alpha} \frac{y^{(2)} (\frac{2\pi}{\Omega})}{R} = 2 \pi \varepsilon_y.\nonumber \ee
Using Eqs.(\ref{Schc}) and (\ref{Solc}) we get respectively

\be
\Delta \varphi_{\rm S} \simeq \frac{6 \pi r_{\rm S}}{R} \;\; [{\rm rad / orbit}]\nonumber \ee
for the relativistic advance of the perihelion and

\be
\Delta \varphi_{\rm Q} \simeq 3 \pi J_2 
\left(\frac{R_{\odot}}{R}\right)^2 \;\; [{\rm rad / orbit}]\nonumber \ee
for the corresponding advance due to the $J_2$ term. These results are in agreement with standard findings in the limit of small eccentricities \cite{Wei} making us confident about the reliability of the solution. The higher order terms are

\ba
x_{1}^{(3)} (t) &=& -2 x_{2}^{(3)}  (t) = -2 x_{3}^{(3)} (t) = \frac{2\varepsilon_x\ell}{\sqrt{3}} \left(1-\cos\Omega t \right),\nonumber\\
y_{1}^{(3)} (t) &=& -\sqrt{3}\varepsilon_y\ell \left( \Omega t-\sin\Omega t \right),\nonumber\\
y_{2}^{(3)} (t) &=& \varepsilon_y\ell \left(-\frac12 + \frac{\sqrt{3}}2 \Omega t-\cos \left(\Omega t - \frac{2 \pi}3 \right) \right),\\
y_{3}^{(3)} (t) &=& \varepsilon_y\ell \left(\frac12 + \frac{\sqrt{3}}2\Omega t+\cos \left(\Omega t - \frac{4 \pi}3 \right) \right),\nonumber\\
z_k^{(3)} (t) &=& -\frac14 \varepsilon_z\ell \left(\Omega t \sin \left(\Omega t - \frac{2 \pi}3 (k-1) \right) - \frac{k-1}{3k-7} \frac{\sqrt{3}}2 \sin\Omega t \right).\nonumber \ea         
         To better show the effects produced by these perturbations we can use the quantities 

\be\label{diffpert}
\Delta L_{jk}^{pert}-\Delta L_{jk}^{kepl}, \ee
where $\Delta L_{kj} = L_{ij}-L_{ik}$ and the arm-lengths $L_{jk}$ are computed using the full perturbed solution (\ref{approx3}) in $\Delta L_{jk}^{pert}$ and including only the Keplerian solutions in $\Delta L_{jk}^{kepl}$. 

\begin{figure}
   \centering
\includegraphics[width=0.8\textwidth]{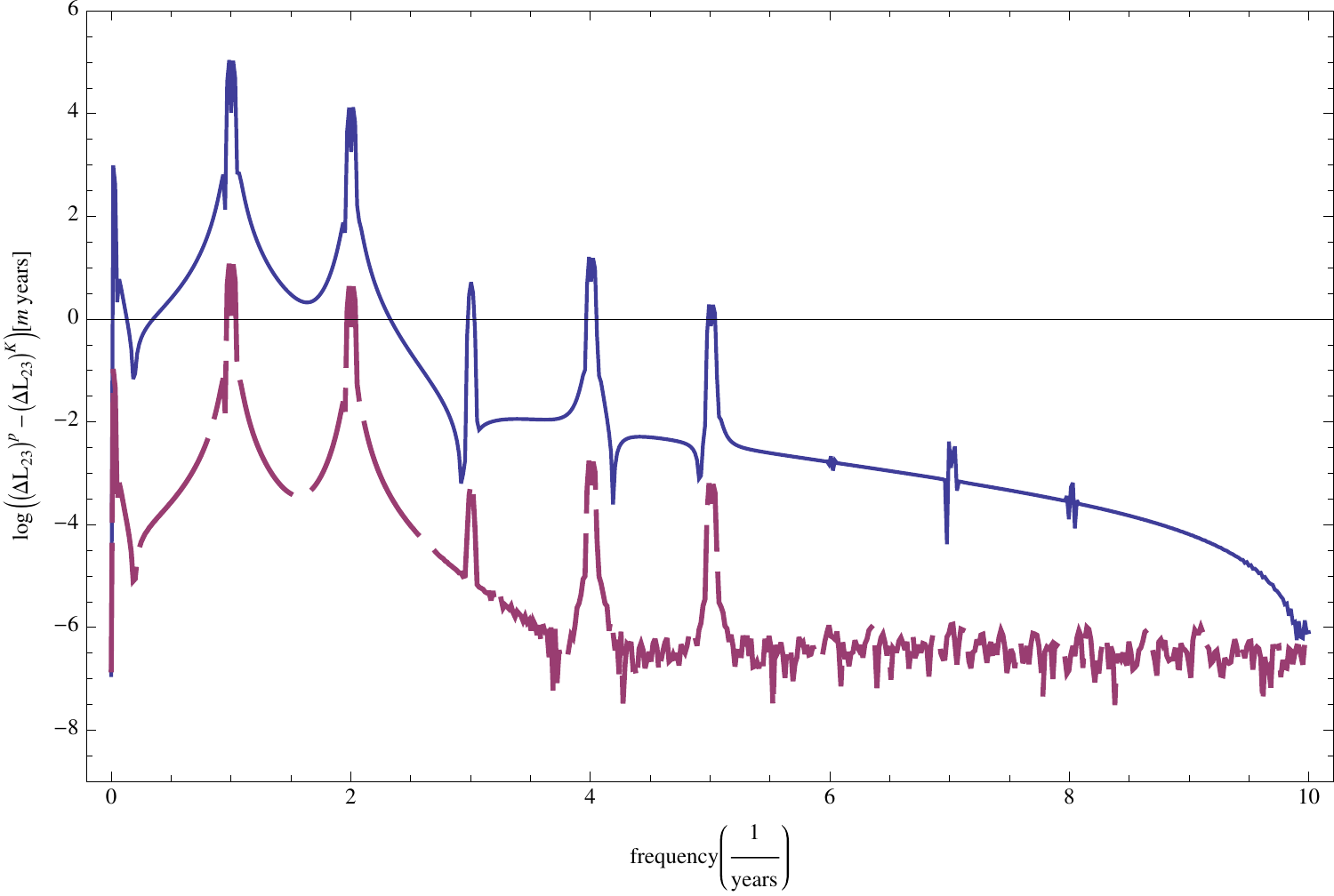}
      \caption{Spectrum of the difference $\Delta L_{32}^{pert}-\Delta L_{32}^{kepl}$: the continuous line represents the effect due to the relativistic precession; the dashed line that due to the Solar quadrupole (with $Q=1$).}
         \label{spr}\end{figure}

In Fig.\ref{pr} we show these quantities  time for a period of 5 years as a consequence of the differential effect due to the relativistic precession: the amplitude of the arm length variations can be as large as 10 km. In Fig.\ref{pq} we show the same quantities 
 due to the differential effect of the Solar quadrupole: the amplitude of the arm length variations reaches 1.5 m times $Q$. The modulus of the Fourier transform of $\Delta L_{32}^{pert}-\Delta L_{32}^{kepl}$ is shown in Fig.\ref{spr}. 
 
 In principle, a precise evaluation of all other effects (Schwarzschild precession included) would allow a measurement of $Q$ (and $J_2$) from the residuals. Indeed, Fig.\ref{spr} shows that the relative content of higher harmonics with respect to the fundamental is larger for the Solar quadrupole precession than for the relativistic one. However, this measurement is hindered by the absolute magnitude of the first effect being orders of magnitude smaller than the latter.

\section{A global stationary potential}

The presence of a stationary global potential in the Solar System can be due to different kinds of diffuse matter distributions. Main candidates are the interplanetary dust and a local dark matter component \cite{cerdo1, cerdo2}. Our concern here is to show that this kind of gravitational source is readily described as an autonomous perturbation and can be analyzed with the methods illustrated above. We consider a very simple model in which a spheroidal distribution can be assumed with rotational symmetry around the (barycentric) $Z$-axis and mirror reflection symmetry with respect to the $X-Y$ plane. Another apparently very crude assumption is that of assuming a linear gradient field as a second order perturbation in our system of equations of motion. This hypothesis is however viable since almost circular orbits are probing the extra field in a small radial range in which it is slowly changing and what can be of practical interest is the possibility of detecting an anisotropy in the field, possibly due to the ellipticity of the spheroidal distribution. The hypothesis is sustained by numerical computations with the interplanetary dust \cite{cerdo2} showing that its perturbing effect is weakly affected by the steepness of the radial component of the density function.

The system of equations that we use is of the form (\ref{eq2_123}) with the perturbation vector $\bv f (\bv x, \bv \varepsilon)$ given by

\be f_x = - \varepsilon_{\rm D} \Omega^{2} x,\quad 
     f_y = - \varepsilon_{\rm D} \Omega^{2} y,\quad
     f_z = -  \beta\varepsilon_{\rm D} \Omega^{2} z.
\ee
Here $\varepsilon_{\rm D}$ is the small parameter associated to the diffuse component and we have assumed a spheroidal distribution parametrized by a shape parameter $\beta$. In the very simple setting sketched above, we may assume

\be
\varepsilon_{\rm D} = \frac{4 \pi G \rho_{\rm D}}{3  \Omega^{2}} , \ee
where $\rho_{\rm D}$ is a suitable average over the density distribution. The solution is still of the form (\ref{approx2}) where in the second order terms we add the contribution due to the solution

\ba\label{seqd}
x_{k \, \rm D}^{(2)} &=& 2  \frac{A'_{k}}{\Omega} + 
 B'_{k} \cos \phi_{k} + C'_{k} \sin \phi_{k} - \frac{5\ell}{4\sqrt{3}} \phi_{k} \sin \phi_{k},\nonumber \\
y_{k \, \rm D}^{(2)} &=& -3 \frac{A'_{k}}{\Omega} \phi_{k} + 2 (C'_{k} \cos \phi_{k} - B'_{k} \sin \phi_{k}) + D'_{k}
- \frac{5\ell}{2\sqrt{3}} \left( \phi_{k} \cos \phi_{k} - \frac35
                        \sin \phi_{k} \right), \\
z_{k \, \rm D}^{(2)} &=& E'_{k} \cos \phi_{k} + F'_{k} \sin \phi_{k} - \frac{\ell}{4} \phi_{k} \sin \phi_{k}\nonumber
  \ea
and the arbitrary constants have a prime to distinguish them from those of the solution for the Earth. With the natural condition of putting to zero both the displacement and its time rate at `zero' (half-mission) time, we get

\begin{align}
A'_{k} &= \frac{\Omega \ell}{\sqrt{3}} \cos \sigma_{k} ,  &   B'_{k} &= -\frac{13 \ell}{8 \sqrt{3}} - \frac{\sqrt{3} \ell}{8} \cos 2 \sigma_{k}, \nonumber \\
C'_{k} &= \ell \frac{3 \sin 2 \sigma_{k} - 10 \sigma_k}{8 \sqrt{3}}, &
D'_{k} &= - \sqrt{3} \ell \sigma_{k} \cos \sigma_{k} + \frac{4 \ell}{\sqrt{3}} \sin \sigma_{k},\\
E'_{k} &= -\frac{\ell}{4} \sin^2 \sigma_{k}  ,&  %\nonumber \\
F'_{k} &= -\frac{\ell}{8} \left( 2 \sigma_{k}  - \sin 2 \sigma_{k} \right).\nonumber
\label{inid}
    \end{align}
The components in (\ref{seqd}), multiplied by $\varepsilon_{\rm D}$ and added to the solution  (\ref{approx}), provide the evolution of the arm-lengths under this perturbing effect. To show the small effects of the diffuse component, Cerdonio et al. \cite{cerdo2} (whose approach is purely numerical) plot the following quantities

\be\label{diffij}
\delta L_{jk} = \Delta L_{jk}^{\rm D} - L_{jk}^{kepl}, \ee
where the arm-lengths $L_{jk \, \rm D}$ are computed using the full perturbed solution and the $L_{jk}$ are those obtained with the optimal Keplerian solution. 

\begin{figure}
   \centering
\includegraphics[width=0.8\textwidth]{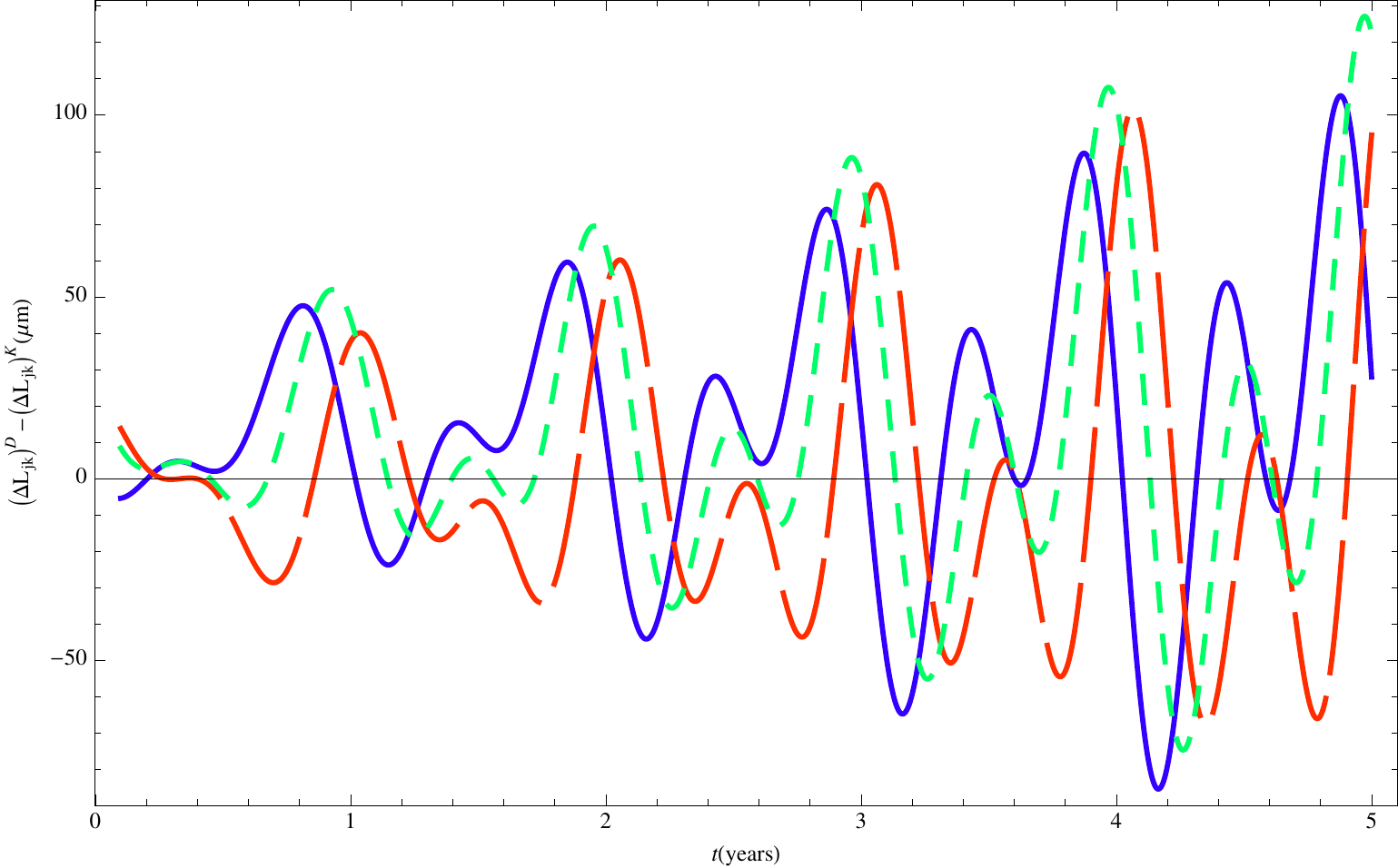}
      \caption{Relative distances $\delta L_{jk}$  time due to a spheroidal ($\beta=\sqrt{3}$) distribution of interplanetary dust, over a time-span of 5 years:  the continuous line is for $jk = 23$, the long dashed line for $jk = 13$ and the short dashed  line for $jk = 12$.}
         \label{Polvere1}\end{figure}

\begin{figure}
   \centering
\includegraphics[width=0.8\textwidth]{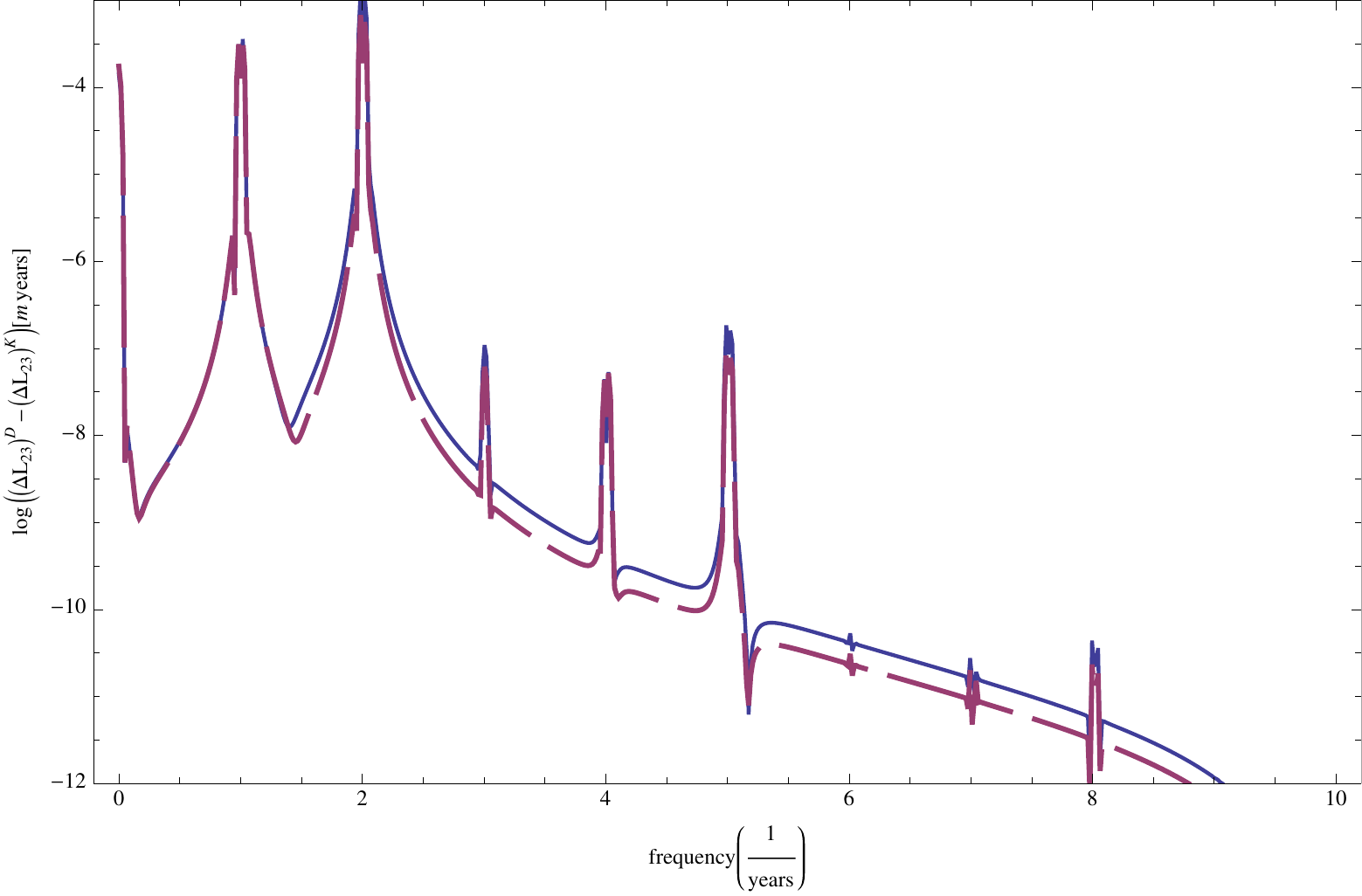}
      \caption{Spectrum of the relative distance $\delta L_{jk}$ for two values of the shape parameter $\beta$: $\beta=\sqrt{3}$ (continuous line) and $\beta=1$ (dashed line).}
         \label{Polvere3}\end{figure}

In Fig.\ref{Polvere1} these relative distances are plotted in the case in which the average density is taken to be $\rho_{\rm D} = 9.6 \times 10^{-20} \, kg \, /m^{3}$ as estimated by models of the meteoritic complex \cite{grun} and a shape parameter $\beta=\sqrt{3}$ (spheroidal distribution): we can compare these plots with those for the $\alpha=0$ case of reference \cite{cerdo2}, corresponding to their most homogeneous distribution. Our solution is in good agreement with that obtained with a numerical integration of the Gauss equations \cite{cerdo2}. In Fig.\ref{Polvere3} we have plotted the Fourier spectrum of two different realizations of the difference $\delta L_{32}$ of Eq.(\ref{diffij}). In ref.\cite{cerdo2} an analogous plot has been obtained by the discrete Fourier analysis of the numerical solution and we can again appreciate the good accuracy of this simple analytical approach. We can see how the solutions suggest the possibility of assessing the shape of the matter distribution through the spectral analysis of the arm lengths variation. However, we have seen in the previous sections that other autonomous perturbations effects leave on these observables marks at the same frequencies but with higher amplitude.


\section{Conclusions and future work}

We have reported on a systematic study of the autonomous perturbations of LISA orbits, namely those 
stationary with respect to the center of the LISA constellation, i.e. constant in the Hill frame. 
This is the simplest step in the study of the perturbations on the LISA orbits.

As in other recent papers \cite{cqg1,cqg2,sw}, we implement analytical techniques based on a `post-epicyclic' approximation in the Hill frame. This work is intended as a preliminary study aimed at gathering an overall picture and more physical insight to the issue. In fact, the analysis of changes of the ideal Keplerian orbits may suggest in principle possible changes in the choice of initial conditions. Accurate, dedicated numerical investigations \cite{hu,pk, bvj} are then required in the assessment of the exact amount of each effect. However, a numerical optimization depends,  in the case of independent orbits, on eighteen parameters and is yet to be fully explored \cite{hu}. A systematic comparison of our analytical approach with these  solutions is a further step that must be undertaken.

A particular emphasis was put on the tidal field of the Earth, assumed to be stationary in the Hill frame. The solution of the equations of motion under the perturbation gives a combination of a reversed parabolic motion in the tangential direction and a drift in the radial direction. This sheds light on possible implications on the choice of initial conditions when the time base-line of the mission is longer than that assumed in previous papers. We propose a more systematic investigation of the phase choice affecting the orientation of the constellation at the moment of closest approach to the Earth that could be of a certain practical relevance in the case it should be necessary to privilege the proper working of only a pair of arms. 

Another relevant class of autonomous perturbations is that given by a global stationary field, associated either to the classical or relativistic corrections to the spherically symmetric Solar potential or to sources like the interplanetary dust or a local dark matter component. The inclusion of a simple linear contribution representative of these fields produces a secular solution that can be used to evaluate physical parameters of the perturbing force. The relativistic precession and the more complex motion due to a Solar quadrupole contribution are most probably within the detection capabilities of LISA. More problematic, with the figures obtained, is a detection of a possible flatness of the perturbing spheroidal component.

The natural step further on this research line is to attempt the analysis of non-autonomous perturbations. The time-varying contributions due to the Moon, the planets and other celestial bodies must be included in the list of relevant perturbing effects. On the basis of the present work, we are confident that the study of a properly modified version of the equations of motion in the form of a forced system of coupled nonlinear oscillators (e.g. that recently presented in ref.\cite{cerdo3}) will provide additional solutions with a degree of accuracy comparable to that presented here.

\section*{Acknowledgments}

We thank Pete Bender, Alfio Bonanno, Roberto De Pietri, Oliver Jennrich, David Lucchesi, Ruggero Stanga and Michele Vallisneri for fruitful discussions.
GP acknowledges partial support from ICRAnet.

\end{document}